\documentclass{article}
\usepackage{amssymb,amsmath,bm}
\begin{document}
\begin{titlepage}
\noindent{\large\textbf{On the relation between the spin and the magnetic moment of the proton}}
\vspace{\baselineskip}
\begin{center}
{Amir~H.~Fatollahi~{\footnote {fath@alzahra.ac.ir}}
\\
Ahmad~Shariati~{\footnote {shariati@mailaps.org}}}\\
Mohammad~Khorrami~{\footnote {mamwad@mailaps.org}}\\
\vspace{\baselineskip}
\textit{ Department of Physics, Alzahra University, Tehran
1993891167, Iran }
\end{center}
\vspace{\baselineskip}
\begin{abstract}
\noindent In the context of the quark model of hadrons the spin and
the magnetic moment of proton can not be taken proportional. This is
in contradiction with the widely used relation between these two properties
of the proton. This apparent difficulty is addressed by the most elementary
notions of the relevant physics. In particular it is emphasized that
the widely used relation is only valid in the lowest orders of perturbation,
in which transitions between different baryons do not occur. For other processes
where such transitions do occur, such as inelastic scattering off the protons,
the quark model relation for the magnetic moment is used to give an estimation
for the amplitude of transition between states with different total spins.
\end{abstract}

\vspace{1cm}


\end{titlepage}
\noindent
It is known that for composite systems, whenever the charge-to-mass ratio
($q_a/m_a$) are not the same for the constituents, the magnetic moment of
the system is not proportional to its total angular momentum (see e.g. p.~187
of \cite{jack}). Further, in atomic physics it is known that, due to different
gyromagnetic factors (the so-called $g$-factors) of the orbital angular momentum
and spin, the magnetic moment of atoms and their total angular momentum are
not proportional (see e.g. p.~301 of \cite{greiner}). It is the purpose of
this note to address a less emphasized similar phenomena for
the composite states of the hadron physics in the context of
the quark model \cite{griff,halzen}.

According to the quark model the hadrons are made up of
quarks. Each quark is specified by its mass, charge, spin,
flavor, and color. The states of baryons with the lowest
masses are symmetric combinations in the space of
``$\mathrm{flavor}\otimes\mathrm{spin}$". For example,
the spin-up proton state is presented by \cite{griff}
\begin{align}\label{1}
|\mathrm{p}\!\uparrow\rangle =\frac{1}{3\sqrt{2}}\Big(
&|\mathrm{udu}-\mathrm{duu}\rangle
|\!\uparrow\downarrow\uparrow - \downarrow\uparrow\uparrow\rangle \nonumber\\
+
&|\mathrm{uud}-\mathrm{udu}\rangle
|\!\uparrow\uparrow\downarrow - \uparrow\downarrow\uparrow\rangle\nonumber\\
+
&|\mathrm{uud}-\mathrm{duu}\rangle
|\!\uparrow\uparrow\downarrow - \downarrow\uparrow\uparrow\rangle\Big),
\end{align}
in which $|\mathrm{u}\rangle$ and $|\mathrm{d}\rangle$ represent the
up and down flavors of quarks, respectively. A similar expression can
be given for the spin-down proton state, $|\mathrm{p}\!\downarrow\rangle$.
It is easy to check that both states $|\mathrm{p}\!\uparrow\rangle$ and
$|\mathrm{p}\!\downarrow\rangle$ are simultaneous eigenstates of the
square and the $z$-component of the total-spin operator
\begin{equation}\label{2}
\mathbf{S}=\mathbf{S}_1+\mathbf{S}_2+\mathbf{S}_3,
\end{equation}
with eigenvalues $\frac{3}{4}\hbar^2$
and $\pm\frac{1}{2}\hbar$, respectively.

In the context of the quark model, the magnetic moment of
a baryon state is defined as the vector sum of
the magnetic moments of its three constituent quarks \cite{griff,halzen}:
\begin{equation}\label{3}
\bm{\mu}=  \bm{\mu}_1+ \bm{\mu}_2+ \bm{\mu}_3,
\end{equation}
in which (SI units)
\begin{equation}\label{4}
\bm{\mu}_a=\frac{Q_a}{m_a}\mathbf{S}_a,\qquad a=1,2,3
\end{equation}
with $Q_a$, $m_a$, and $\mathbf{S}_a$ representing the electric charge, the constituent
mass, and the spin operator of the $a\;\!$'th quark, respectively.
As is evident, the two vectors (\ref{2}) and (\ref{3}), even after equating the
constituent masses $m_a$'s, are not parallel. Further,
one can easily check that neither $|\mathrm{p}\!\uparrow\rangle$ nor
$|\mathrm{p}\!\downarrow\rangle$ are eigenstates of the
$z$-component of the vector (\ref{3}). The reason simply is that
the electric charges of up and down flavors are different.
The commutation relations
\begin{align}
\label{5}
[\mu_i,S_i]&=0,  \\
\label{6}
[\mu_i, \mathbf{S}\cdot\mathbf{S}]&\neq 0,\quad i=x,y,z
\end{align}
mean that, for example, the $\mu_z$ and $S_z$ can be determined
simultaneously, but not in the basis in which the total-spin is given.
In fact, in the context of the quark model it is known that
the matrix element of the magnetic moment operator between
different hadronic states, the so-called transition magnetic moment,
can give an estimation for the amplitude of the transition between
states with different total spins as well. For example, the matrix 
element $\langle\mathrm{p}|\mu_z|\Delta^+\rangle\neq 0$ can be 
used to estimate the amplitude at the corresponding resonance 
\cite{pais} (see \cite{fay,khan} for a presentation in the 
nowadays notation).

The simple observations above should be contrasted with the widely used
expression, that the relation between the spin and magnetic moment
of proton is given by \cite{griff,halzen,messiah,merz,leader}
\begin{equation}\label{7}
\bm{\mu}_\mathrm{p} =g_\mathrm{p}\frac{e}{2\,M_\mathrm{p}}\,\mathbf{S}_\mathrm{p},
\qquad g_\mathrm{p}=5.59
\end{equation}
saying that
\begin{itemize}
\item[1)] the spin and magnetic moment of the proton are parallel,
\item[2)] the total spin squared ($\mathbf{S}\cdot\mathbf{S}$),
as well as the $z$-components of the total-spin and
the total magnetic moment ($S_z$ and $\mu_z$, respectively)
can be determined simultaneously.
\end{itemize}
So the above mentioned transitions between states with different spins
would have been forbidden by relations like (\ref{7}).
It is the purpose of this work to clarify these within
the context of the quark model, and to explore the domain of
validity of (\ref{7}). In particular, it is emphasized that
the relation (\ref{7}) should be interpreted as a relation between
the matrix elements of the spin and the magnetic moment in the lowest order of
perturbation, where transition between different baryon states does not occur.
Hence if the energy due to the coupling of the magnetic moment with
the external field is so high that transition from one baryon state to
another one is possible, the relation (\ref{7}) is not valid anymore.

Here only baryons with the quark content ``uud" (spin-1/2 $\mathrm{p}$
and spin-3/2 $\Delta^+$ \cite{griff,halzen}) are studied. Though it should be
emphasized that a more or less similar reasoning shows that the present analysis
applies to other hadronic states as well.

As an example, the $\Delta^+$ state with $s_z=\frac{3}{2}\hbar$ is given by \cite{griff}
\begin{equation}\label{8}
\left|\Delta^+, +\frac{3}{2}\right\rangle =
\frac{1}{\sqrt{3}}\,|\mathrm{uud+udu+duu}\rangle\,|\!\uparrow\uparrow\uparrow\rangle
\end{equation}
One can give similar expressions for $\Delta^+$ states with
$s_z=\frac{1}{2}\hbar,-\frac{1}{2}\hbar,-\frac{3}{2}\hbar$.
If all other interactions had been turned off, the two $\mathrm{p}$ states and
the four $\Delta^+$ states would have been the eigenstates of the mass operator
\begin{equation}\label{9}
M_0=m_1+m_2+m_3
\end{equation}
with the eigenvalue $2\, m_\mathrm{u}+m_\mathrm{d}$. So the proton and $\Delta^+$
would have equal masses. In the context of the quark model, a strong version
of the hyperfine effect is responsible for the mass splitting between these
two baryons \cite{griff}. In the case of our interest, assuming
$m_\mathrm{u}=m_\mathrm{d}=m$, we have
\begin{equation}\label{10}
M'_0=3\, m+ \frac{A}{m^2}\,
(\mathbf{S}_1\cdot \mathbf{S}_2+\mathbf{S}_2 \cdot \mathbf{S}_3+\mathbf{S}_1 \cdot \mathbf{S}_3)
\end{equation}
in which the free constant $A$ is tuned such that the best fit to the mass data is
achieved \cite{griff}. Since the above hyperfine term commutes with the square of
total spin, $\mathbf{S}\cdot\mathbf{S}$, the mass degeneracy between the six states
of the proton and $\Delta^+$ is reduced to a degeneracy between the two states
of the proton and a degeneracy between the four states of $\Delta^+$:
\begin{align}\label{11}
M_{\mathrm{p}}&=938~\frac{\mathrm{MeV}}{c^2},\nonumber\\
M_{\Delta^+}&=1232~\frac{\mathrm{MeV}}{c^2}.
\end{align}
The unperturbed rest mass operator in the sub-space of
the ``uud" quark content and in the basis of the simultaneous
eigenvectors of $\mathbf{S}\cdot\mathbf{S}$ and $S_z$ is in the form
\begin{equation}\label{12}
M'_0=\begin{pmatrix}
M_\mathrm{p} & 0 & 0 & 0 & 0 & 0 \\
0 & M_\mathrm{p} & 0 & 0 & 0 & 0 \\
0 & 0 & M_{\Delta^+} & 0 & 0 & 0 \\
0 & 0 & 0 & M_{\Delta^+} & 0 & 0 \\
0 & 0 & 0 & 0 & M_{\Delta^+} & 0 \\
0 & 0 & 0 & 0 & 0 & M_{\Delta^+}
\end{pmatrix}.
\end{equation}

Let us now address the issue of the magnetic moment. The magnetic moment
of particles are usually measured by monitoring the behavior of them in
an external magnetic field $\mathbf{B}(\mathbf{r},t)$. Such a field
results in an interaction term
\begin{equation}\label{13}
\delta H=-\bm{\mu}\cdot\mathbf{B}(\mathbf{r},t).
\end{equation}
Common examples of such experiments are the Stern-Gerlach or the spin resonance
set-ups. As mentioned before, baryonic states of definite total spin squared
for which the constituent flavors have different electric charges, are
not eigenstates of a component of the total magnetic moment. This means that
the perturbation Hamiltonian (\ref{13}) has off diagonal elements. Subsequently,
one should expect that the transitions between different ``uud" states
are possible. In other words, for the matrix elements which are responsible
for the previously mentioned proton-$\Delta$ transition
\cite{pais,fay,khan}, one can directly show that
\begin{align}\label{14}
\langle \mathrm{p},s_z| \bm{\mu}\cdot\mathbf{B} | \Delta^+, s_z\rangle \neq 0,
\qquad& \mathbf{B}\parallel\mathbf{\hat{z}}, \\ \label{15}
\langle \mathrm{p},s_z| \bm{\mu}\cdot\mathbf{B} | \Delta^+, s'_z\rangle \neq 0,
\qquad& \mathbf{B}\perp\mathbf{\hat{z}}~\;\&\;~s_z\neq s'_z,
\end{align}
in which $\bm{\mu}$ is given by (\ref{3}). One should notice that by
using $\bm{\mu}_\mathrm{p}$ of (\ref{7}) instead of (\ref{3}) in the above,
the matrix elements which are responsible for the transition
$\mathrm{p}\leftrightarrows \Delta^+$ would vanish, simply
due to the fact that proton belongs to the spin-1/2 subspace
while $\Delta^+$ belongs to the spin-3/2 one.
In the perturbative regime, $|\mu\,B|\ll(M_{\Delta^+}-M_{\mathrm{p}})\,c^2$, however,
up to first order in perturbation only those matrix elements of
 $\delta H$ contribute which correspond to states inside each of
the degenerate blocks of the original Hamiltonian \cite{messiah,merz}.
For the proton block, it can be directly shown that
\begin{equation}\label{16}
\langle\mathrm{p},s_z|\bm{\mu}|\mathrm{p},s'_z\rangle=\frac{2}{\hbar}\,\mu_\mathrm{p}\,
\langle\mathrm{p},s_z|\mathbf{S}|\mathrm{p},s'_z\rangle,
\end{equation}
where
\begin{equation}\label{17}
\mu_\mathrm{p}:=\langle\mathrm{p}\!\uparrow\! |\mu_z|\mathrm{p}\!\uparrow\rangle.
\end{equation}
Equation (\ref{17}) is precisely the same that is given as the strength of
the magnetic moment of proton in the context of the quark model of hadrons \cite{griff}.
It is seen that, as far as the lowest order of perturbation is concerned, one can
employ (\ref{7}), in which the spin and the magnetic moment of proton are parallel.

Let us elaborate it a little more. In general the magnetic moment has
two origins, the spin and the orbital angular momentum. Regarding
the proton, the contribution from the latter does not exist, as
the quarks inside the proton are in a state of zero orbital angular momentum.
Hence (\ref{3}) and (\ref{4}) follow. It could then be expected that the
magnetic moment of the proton should be a c-number times the spin of the
proton, as the proton is spherically symmetric from the point of view of
the orbital motion of its constituent quarks. This would be true if
the magnetic moment and the spin were vectors with c-number components,
and in fact it is true for the expectation value of the magnetic moment
and the spin, as stated in (\ref{16}). Equation (\ref{16}) is even more than
that. Not only the expectation values, but the general matrix elements of
the magnetic moment are and the spin are proportional by
a single proportionality constant, provided one is restricted to the
subspace corresponding the proton. The point is that the magnetic moment
operator is not equal to a c-number times the spin operator.
The magnetic moment operator is a c-number times the spin operator,
in the subspace of proton, and another c-number times the spin operator
in the subspace of $\Delta^+$. Moreover, the magnetic moment operator
does have nonzero matrix elements between these two subspaces, while
the spin operator is not so:
\begin{align}\label{18}
\langle\mathrm{p},s_z|\bm{\mu}|\mathrm{p},s'_z\rangle&=c_\mathrm{p}\,
\langle\mathrm{p},s_z|\mathbf{S}|\mathrm{p},s'_z\rangle,\nonumber\\
\langle\Delta^+,s_z|\bm{\mu}|\Delta^+,s'_z\rangle&=c_{\Delta^+}\,
\langle\Delta^+,s_z|\mathbf{S}|\Delta^+,s'_z\rangle,\nonumber\\
\langle\Delta^+,s_z|\bm{\mu}|\mathrm{p},s'_z\rangle&\ne0,\quad\hbox{in general}.
\end{align}
This is a manifestation of the Wigner-Eckart theorem.

As mentioned before, turning on the interaction in principle mixes the two blocks
corresponding to the proton and $\Delta^+$. In order that this mixing be
significant, the matrix elements of the perturbation Hamiltonian
should be comparable with the difference between the eigenvalues of the
unperturbed Hamiltonian. Let us have an estimate for the threshold value
of the magnetic field, corresponding to such a mixing
\begin{equation}\label{19}
(\mu_\mathrm{p}\,B)\sim(M_{\Delta^+}-M_{\mathrm{p}})\,c^2,
\end{equation}
resulting in
\begin{equation}\label{20}
B\sim3\times10^{15}\;\mathrm{T}.
\end{equation}
Even at the surface of a magnetar, the magnetic field reaches only up to
$10^{11}$~T. So, one can use the relation (\ref{7}) with extreme safety in almost all
practical cases. But why almost?

As mentioned before, in the situations in which transitions between the baryon states are
possible, the lowest orders of perturbation, by which the relation (\ref{7}) is justified,
are insufficient. In fact, in the processes of scattering off the protons we are faced with
this situation. Let us make an estimation on the possibility of generation of the threshold
value mentioned above in the scattering of electrons off the rest protons.
The magnetic field of an ultra-relativistic electron at transverse distance
$b$ is given by \cite{jack}
\begin{equation}\label{21}
B_\mathrm{e}\sim\frac{\mu_0\,c}{4\,\pi}\gamma\,\frac{e}{b^2},
\end{equation}
where $\gamma$ is the Lorentz factor. The corresponding de~Broglie
wavelength is
\begin{align}\label{22}
\lambda&=\frac{h}{\gamma\,m\,c},\nonumber\\
&\sim\frac{2\times 10^{-12}}{\gamma}\;\mathrm{m},
\end{align}
showing that for an electron of the Lorentz factor 2000
(energy equal to 1~GeV), this wavelength is of the order $10^{-15}$~m.
So such an electron can be close to a proton down to such a distance.
It is then seen that a threshold field can be produced by such an
electron at a distance of the order $10^{-15}$~m. In fact, in
inelastic scattering experiments, transitions of the type
$\mathrm{e}\,\mathrm{p}\to \mathrm{e}\,\Delta \to \mathrm{e}\,\mathrm{p}\,\pi^0$
are considered as the ones with the modest amount of the energy transfer between
the incoming electron and the proton \cite{halzen}. 

If the magnetic field is so large that (\ref{19}) holds, then the mass matrix
would be no longer like (\ref{12}). It would have non-diagonal terms
(in the basis consisting of $\mathrm{p}$ and $\Delta^+$ states). So
$\mathrm{p}$ and $\Delta^+$ would no longer be eigenstates of mass.
The mass eigenstates would be mixtures of $\mathrm{p}$ and $\Delta^+$.
One would then expect an oscillation between $\mathrm{p}$ and
$\Delta^+$. So a very high magnetic field (coming from whatever
source, an inelastic scattering for example) could induce an
oscillation between the $\mathrm{p}$ and $\Delta^+$ states.
The off-diagonal matrix elements which are responsible for
such an oscillation are easily calculated. For example, consider 
a magnetic field in the $z$ direction. One has,
\begin{equation}\label{23}
\left\langle\Delta^+,s_z=\frac{\hbar}{2}\right|(-\mu_z\,B)\left|\mathrm{p},s_z=\frac{\hbar}{2}\right\rangle
=-\frac{\sqrt{2}\,B\,\hbar}{3}\,\left(\frac{Q_\mathrm{u}}{m_\mathrm{u}}-
\frac{Q_\mathrm{d}}{m_\mathrm{d}}\right),
\end{equation}
which vanishes if and only if the charge to mass ratio is the same for the
up and down quarks. So the oscillation occurs, if and only if the charge to mass
ratio is different for the up and down quarks, or equivalently if and only if
the magnetic moment is not parallel to the spin.
\\[\baselineskip]
\textbf{Acknowledgement}:
This work is supported by the Research Council of the Alzahra University.

\end{document}